# The Handheld and Hand Powered Homopolar Generator

Keith Zengel[1]
1. Harvard University, Cambridge, MA

The homopolar or unipolar generator, which is sometimes referred to as a Faraday Paradox, is one of my favorite lab topics to teach. At the end of a lab on induced emfs, I like to propose the setup as a separate puzzler to the students: If you connect the leads of a current sensor to the center and edge of a conducting magnet, as shown in Fig. 1, then rotate the magnet, do you expect to register a current? Why? With the flux rule and Lenz's law fresh in their minds, most students will answer that because there is no changing flux (rotating the magnet does not affect the affect the amount of magnet enclosed by the leads), so there is no induced emf. Students are typically surprised to find that they do indeed measure a current that depends on how quickly and which direction they rotate the magnet. Their confusion usually intensifies when they realize that the Lorentz force law *does* predict an emf, which seems inconsistent with the flux rule.

They are not alone in their confusion. Richard Feynman included this example as one of his so-called "exceptions to the flux rule."[1] Faraday himself thought the phenomena "striking," and after endearingly deciding in 1831 that the underlying law is "very simple, although rather difficult to express,"[2] we find that twenty years later he was still working out the details.[3]

Most demonstrations of this phenomenon have included relatively elaborate setups that require leads connected to a magnet rotating at high angular velocities.[4] While effective, these apparatuses are often expensive, complicated, costly, or time intensive to build, all of which may deter instructors and intimidate students.

An alternative setup is to use a strong neodymium magnet and a Vernier current sensor.[5] The DX88 magnets offered by K&J Magnetics[6] are 1.5'' in diameter and 0.5'' thick, have an axial field that varies from 0.36 T to 0.46 T, and sell for $23. The emf induced by connecting the leads to the center and edge of the magnet is given by $E = w*B*r^2/2$. The Vernier current sensor has a resistor of 0.1 Ohms, so for a $w$ on the order of one, we can expect currents on the mA level. Students can hold the leads in place with one hand and rotate the magnet with the other to explore and contemplate the currents that they measure in LoggerPro. Alternatively, students may use an oscilloscope to measure the induced emf.

My favorite resolution to the paradox is given by Scanlon, Henriksen, and Allen.[7] They encourage readers to begin by calculating the emf predicted by taking a line integral of the Lorentz Force acting on electrons in the surface of the magnet. This emf can plugged into the flux rule to solve for the flux that is supposedly changing with time. Students will find that the area through which the flux is supposedly changing is the area enclosed by the closed loop of the leads and an arbitrary path that connects the two leads across the surface of the magnet. I like this method because it teaches students a strategy for dealing with situations where fundamental laws apparently disagree: solve the problem you can and work backwards to the problem you can't.

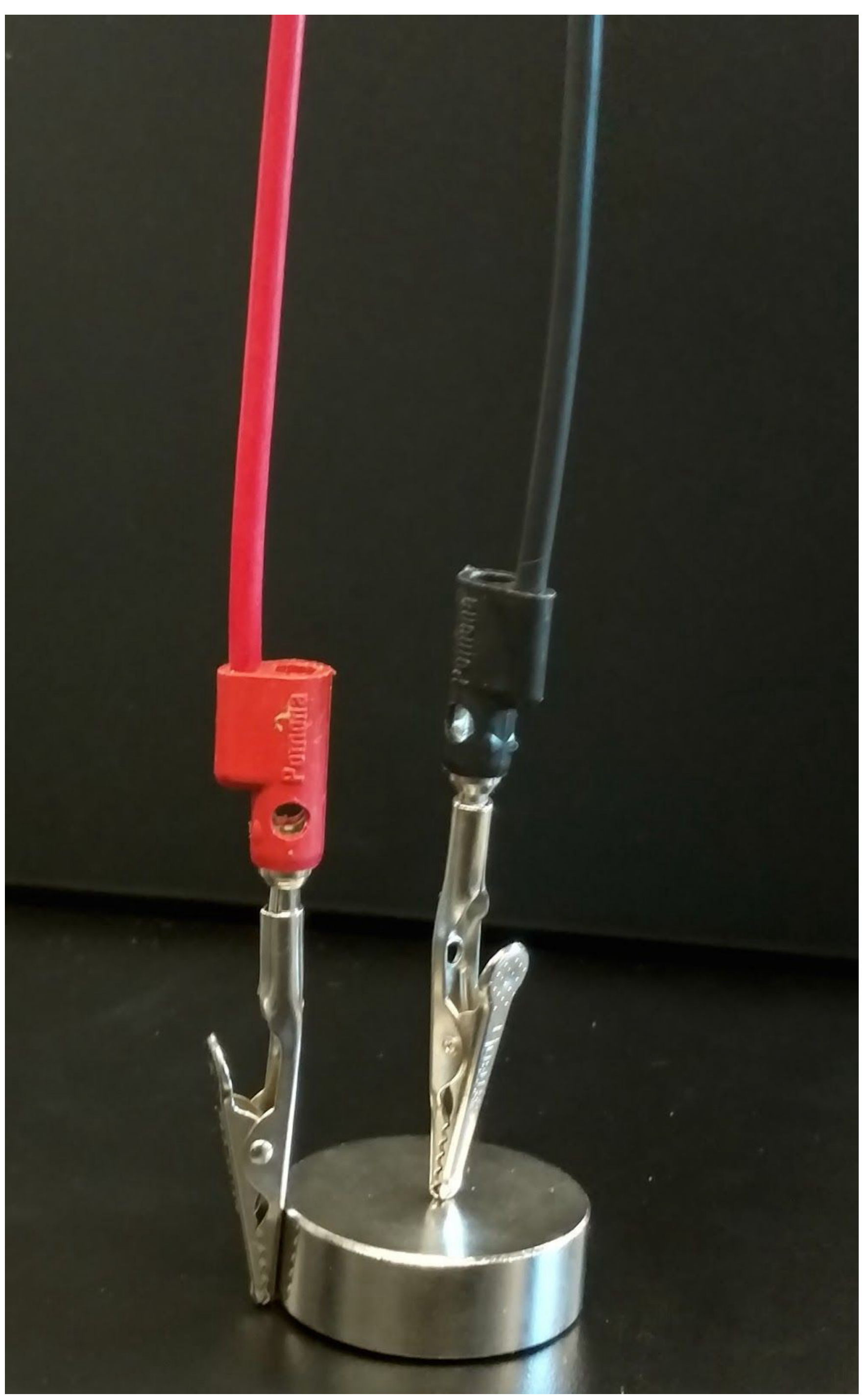

**Fig. 1** Contact locations for the leads on the magnet. The leads are connected across a Vernier current sensor (not shown).

[1] R. P. Feynman, R. B. Leighton, and M. Sands, *The Feynman Lectures on Physics, Vol. II* (Addison-Wesley, Reading, MA, 1964), Sec. 17-2.
[2] Michael Faraday, “V. Experimental Researches in Electricity” and “VI. The Bakerian Lecture–Experimental Researches in Electricity–Second Series,” *Phil. Trans. R. Soc. Lond.* 122,